\documentclass[prd,amssymb,aps,twocolumn,epsfig]{revtex4}
\usepackage{graphicx}
\usepackage{dcolumn}   
\usepackage{amsmath,amssymb}
\usepackage{bm}
\usepackage[colorlinks,
              linkcolor=blue,
              citecolor=blue,
              urlcolor=blue,
              anchorcolor=blue,
              dvipdfm]{hyperref}

\def\be{\begin{equation}}       \def\ee{\end{equation}}
\def\bea{\begin{eqnarray}}      \def\eea{\end{eqnarray}}
\def\ba{\begin{array} }
\def\ea{\end{array} }
\def\bc{\begin{center}}
\def\ec{\end{center}}
\def\bnum{\begin{enumerate} }
\def\enum{\end{enumerate}}

\def\=>{\Rightarrow}
\def\>{\rightarrow}

\date{\today}

\begin{document}

\title{Further Investigation on Model-Independent Probe of Heavy Neutral
Higgs Bosons at the LHC Run 2}

\author{
Yu-Ping Kuang$^{1,2;1}$\email{ypkuang@mail.tsinghua.edu.cn}\quad
Hong-Yu Ren$^{1;2}$\email{renhy10@mails.tsinghua.edu.cn}\quad
Ling-Hao Xia$^{1;3}$\email{xlh10@mails.tsinghua.edu.cn} }

\affiliation{$^1$ Department of Physics, Tsinghua University,
Beijing, 100084, China}

\affiliation{$^2$ Center for High Energy Physics, Tsinghua
University, Beijing, 100084, China}

\begin{abstract}
In our previous paper \cite{KRX-PRD2014}, we provided general
effective Higgs interactions for the lightest Higgs boson $h$
(SM-like) and a heavier neutral Higgs boson $H$ based on the
effective Lagrangian formulation up to the dim-6 interactions, and
then proposed two sensitive processes for probing $H$. We showed
in several examples that the resonance peak of $H$ and its dim-6
effective coupling constants (ECC) can be detected at the LHC Run
2 with reasonable integrated luminosity. In this paper, we further
perform a more thorough study of the most sensitive process,
$pp\to VH^\ast\to VVV$, on the information about the relations
between the $1\sigma,\,3\sigma,\,5\sigma$ statistical significance
and the corresponding ranges of the Higgs ECC for an integrated
luminosity of 100 fb$^{-1}$. These results have two useful
applications in the LHC Run 2: (A) realizing the experimental
determination of the ECC in the dim-6 interactions if $H$ is found
and,  (B) obtaining the theoretical exclusion bounds if $H$ is not
found. Some alternative processes sensitive for certain ranges of
the ECC are also analyzed.\\\\

\null\noindent{PACS numbers: 14.80.Ec, 12.60.Fr, 12.15.-y}
\end{abstract}

\null\noindent{\null\hspace{5cm}TUHEP-TH-151839}

\maketitle

\section{Introduction}

After the discovery of the 125 GeV Higgs in 2012 at the CERN LHC
\cite{ATLAS-CMS12}, the ATLAS and CMS collaborations have measured
its couplings to other
particles\cite{ATLAS-higgs-measurement}\cite{CMS-higgs-measurement}.
So far, to the present experimental precision, they turn out to be
all consistent with the standard model (SM) predictions. However,
it does not mean that the SM is the final theory of fundamental
interactions since it has several shortcomings, such as
unnaturalness\cite{unnaturalness}, triviality\cite{triviality},
vacuum instability\cite {vacuum-instability} and its lack of a
suitable dark matter candidate. Searching for new physics beyond
the SM is still the main task in the TeV scale particle physics.
So far, there is no evidence of the well-known new physics models
such as supersymmetry, large extra dimensions, etc.

We know that most new physics models contain several Higgs bosons,
and the lightest one may behave as (or very close to) the SM Higgs
boson, while the masses of other heavy Higgs are usually in the
few hundred GeV to 1 TeV range. Therefore, the discovered 125 GeV
Higgs boson may actually be the lightest Higgs boson in a new
physics model. So that searching for a heavier Higgs boson may be
a feasible way to find the evidence of new physics. Heavy Higgs
bosons in several most popular models such as the minimal
supersymmetric extension of the standard model (MSSM) and the
two-Higgs-doublet model (2HDM) \cite{ATLAS-PRD89} were searched
for during the LHC Run 1, but no positive evidence has been found.
Therefore, a model-independent probe of the neutral heavy Higgs
bosons is a more efficient way of doing it. In our previous paper
\cite{KRX-PRD2014}, we provided general effective Higgs
interactions for the lightest Higgs boson $h$ (SM-like) and a
heavier neutral Higgs boson $H$ based on the effective Lagrangian
formulation up to the dim-6 interactions, and then we proposed two
sensitive processes, namely the weak-boson scattering $VV\to VV$
(WBS) and $pp\to VH^\ast\to VVV$ (VH$^\ast$), where $V=W,Z$, for
probing $H$. We showed in several examples that the resonance peak
of $H$ and its dim-6 effective coupling constants (ECC) can be
detected at the LHC Run 2 with reasonable integrated luminosity.
Experimentally, the CMS collaboration performs a more general
search, which gives the exclusion limit for a neutral heavy Higgs
boson with the SM couplings up to an overall factor
$C'$\cite{CMS-heavy-higgs}.

In this paper, as in Ref.\,\cite{KRX-PRD2014}, we consider an
arbitrary new physics theory containing more than one Higgs fields
$\Phi_1$, $\Phi_2,\ldots$ without specifying the number of
$\Phi^{}_i$ and their representations. Their interaction potential
$V(\Phi_1,\Phi_2,\ldots)$ may, in general, cause mixing between
the Higgs fields, and form a set of mass eigenstates. We denote
the lightest mass eigenstate by $\Phi_h$, and the second lightest
one by $\Phi_H$. 
The neutral Higgs bosons in $\Phi^{}_h$ and $\Phi^{}_H$ will be
denoted by $h$ and $H$, respectively. Here we identify $h$ with
the discovered 125 GeV Higgs boson.

In the language of effective Lagrangian, we expand the effective
interactions up to the dim-6 terms. Since $h$ is identified with
the discovered 125 GeV SM-like Higgs boson with vanishing dim-6
interactions. For $H$, the effective interactions can be expressed
by
\begin{equation}                       
\label{expantation}
  \mathcal{L}=\mathcal{L}^{(4)}+\mathcal{L}^{(6)}.
\end{equation}

Since $\Phi^{}_H$ is a mixture of the original Higgs Fields
$\Phi^{}_1,\Phi^{}_2,\ldots$, the gauge coupling $g^{}_H$ and
vacuum expectation value (VEV) $v^{}_H$ of $H$ may be different
from the original coupling $g$ and the VEV $v$. We define
\begin{equation}                  
\rho^{}_H\equiv \frac{g^2_Hv^{}_H}{g^2v} \label{rho_H}
\end{equation}
to reflect the mixing effect. the dim-4 term in
Eq.\,(\ref{expantation}) can then be expressed as:
\begin{equation}                   
\label{dim-4_HVV}
\begin{array}{ll}
\mathcal{L}^{(4)}_{HWW} &= gM_{W}\rho_{H}HW_{\mu}W^{\mu},\\
\mathcal{L}^{(4)}_{HZZ} &= \displaystyle\frac{gM_{W}\rho_H}{2c^2}HZ_{\mu}Z^{\mu}.\\
\end{array}
\end{equation}
where $c\equiv \cos \theta^{}_W$.

The dim-6 interactions between $H$ and gauge bosons can be written
through effective Lagrangian as:
\begin{equation}                 
\label{dim-6_operators} \mathcal{L}_{HVV}^{(6)}=\sum_n
\frac{f_n}{\Lambda^2} \mathcal{O}_n
\end{equation}
where $\Lambda$ is the scale under which the effective Lagrangian
works. Here we take $\Lambda=3$ TeV which is consistent with the
theoretical argument $\Lambda\sim 4\pi v$ and with the present LHC
results on heavy particle searches. $\mathcal{O}_n$ are dim-6
operators composed of $H$ and the $SU(2)^{}_L\times U(1)$ gauge
fields with extra derivatives \cite{Hagiwara,buchmuller,G-G}. The
dim-6 $HWW$ and $HZZ$ interactions relevant to this study are
\begin{eqnarray}                              
&&\hspace{-0.4cm}{\cal L}^{(6)}_{HZZ}=
g^{(1)}_{HZZ}Z_{\mu\nu}Z^\mu\partial^\nu H~~
+g^{(2)}_{HZZ}HZ_{\mu\nu}Z^{\mu\nu}\nonumber\\
&&\null\hspace{-0.4cm}{\cal L}^{(6)}_{HWW}=g^{(1)}_{HWW}(W^+_{\mu\nu} W^{-\mu}\partial^\nu H+{\rm h.c.})\nonumber\\
&&+g^{(2)}_{HWW}HW^+_{\mu\nu}W^{-\mu\nu}, \label{L^6_HVV}
\end{eqnarray}
in which 
\begin{eqnarray}                           
&&\null\hspace{-0.4cm}g^{(1)}_{HZZ}=gM^{}_W\rho^{}_H\frac{c^2f_W+s^2f_B}{2c^2\Lambda^2},
\nonumber\\
&&\null\hspace{-0.4cm}
g^{(2)}_{HZZ}=-gM^{}_W\rho^{}_H\frac{s^4f_{BB}
+c^4f_{WW}}{2c^2\Lambda^2},\nonumber\\
&&\null\hspace{-0.4cm}g^{(1)}_{HWW}=gM^{}_W\rho^{}_H\frac{f_W}{2\Lambda^2},~~
g^{(2)}_{HWW}=-gM^{}_W\rho^{}_H\frac{f^{}_{WW}}{\Lambda^2},~~~~~~
\label{g} 
\end{eqnarray}
where $c\equiv \cos\theta^{}_W,s\equiv\sin\theta^{}_W$. Because of
the smallness of $s^2$, Eq.\,(\ref{g}) is mainly described by two
effective coupling constants (ECC) $\rho^{}_H f^{}_W/\Lambda^2$
and $\rho^{}_H f^{}_{WW}/\Lambda^2$ \cite{KRX-PRD2014}.

In the interactions between $H$ and fermions, the main relevant
one is the $H\bar t t$ interaction. It has been shown that, up to
dim-6 terms, the $H\bar t t$ interaction can be expressed as
\begin{equation}                          
\mathcal{L}_{Ht\bar{t}}=y^{H}_t\,\bar{t}^{}_L\Phi^{}_H t^{}_R+{\rm
h.c.}\equiv C^{}_t\,y^{SM}_f\,\bar{t}^{}_L
          \Phi^{}_H t^{}_R+{\rm h.c.},\\
\end{equation}
where $C^{}_t$ is a parameter reflecting the deviation from the SM
Yukawa coupling constant.

Now we have altogether five parameters, namely the mass of the
heavy Higgs boson $M^{}_H$, the anomalous Yukawa coupling factor
$C_t$, the anomalous gauge coupling constant $\rho^{}_H$ in the
dim-4 HVV interaction, and the anomalous coupling constants
$f^{}_W$ and $f^{}_{WW}$ in the dim-6 HVV interactions. They
characterize the heavy neutral Higgs boson $H$
model-independently. In our study, we take $M^{}_H=$ 400 GeV, 500
GeV, and 800 GeV to represent three ranges of $M^{}_H$.

In Ref.\,\cite{KRX-PRD2014}, we pointed out, via several examples,
that VH$^\ast$ and WBS are sensitive processes for discovering $H$
and detecting its ECC $\rho^{}_H f^{}_W/\Lambda^2$ and $\rho^{}_H
f^{}_{WW}/\Lambda^2$. In this paper, we shall give a more thorough
analysis on the relations between the
$1\sigma,\,3\sigma,\,5\sigma$ statistical significance and the
corresponding ranges of the four ECC for the most sensitive
process VH$^\ast$ for an integrated luminosity of 100 fb$^{-1}$.
If signal of the neutral heavy Higgs boson $H$ is detected at the
$3\sigma$ (evidence) or $5\sigma$ level (discovery) level, this
analysis can provide the specific way of realizing the
experimental determination of $\rho^{}_H f^{}_W/\Lambda^2$ and
$\rho^{}_H f^{}_{WW}/\Lambda^2$. If no signal of $H$ is seen, the
$1\sigma$ analysis can provide the theoretical exclusion bounds
\cite{footnote} on the ECC. In certain ECC ranges, the
conventional on-shell production of $H$ via gluon fusion (GF) and
vector-boson fusion (VBF) may also help to discover $H$. We shall
also present the corresponding analysis on these processes.

This paper is organized as follows. First, we give a more detailed
study on the exclusion bounds on the ECC from the LHC Run 1 data
(EB) and the unitarity bound (UB) from the requirement of
unitarity of the $S$ matrix element in Sec.\,2. We first consider
only the dim-4 interactions, and then, without losing generality,
we take into account of the dim-6 interactions by taking certain
sample values of $C^{}_t$ and $\rho^{}_H$ to provide the
two-dimensional plots on the exclusion bounds in the $\rho^{}_H
f^{}_W/\Lambda^2$-$\rho^{}_H f^{}_{WW}/\Lambda^2$ plane for
various values of $M^{}_H$. In Sec.\,3, we provide the analysis on
the information about the relation between the
$1\sigma,3\sigma,5\sigma$ statistical significance and the ranges
of the four ECC for the most sensitive process VH$^\ast$ at the
LHC Run 2 taking account of the present bounds given in Sec.\,2.
In Section 4, we give the
results for the GF and VBF processes. 
Sec.\,4 is a discussion on the exclusion bounds if the signal of
$H$ is not seen at the LHC Run 2.

\section{Exclusion bounds from the LHC Run 1 data and the unitary bound}

In Ref.\,\cite{KRX-PRD2014}, we have studied the exclusion bounds
from the requirement of the unitarity of the $S$ matrix elements
and from the CMS data on excluding the SM-like Higgs boson with
mass from 100 GeV to 1 TeV \cite{CMS-PAS-HIG-13-002} only for
several examples. Now we make a more thorough study of the bounds.

Since the on-shell GF Higgs production process in the LHC Run 1 is
not sensitive to dim-6 interactions, we first study the exclusion
bound without taking account of the dim-6 interactions. Then there
are only two parameters $C^{}_t$ and $\rho^{}_H$ left.

\begin{figure}[h]
\begin{center}
\begin{minipage}{0.45\textwidth}
\centering
\includegraphics[width=0.6\textwidth]{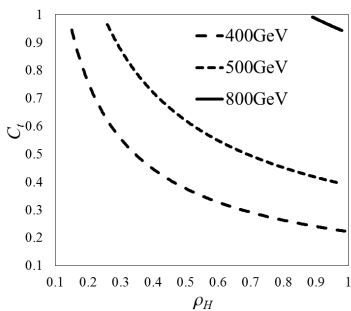}     
\vspace{5pt} \caption{EB for the cases of $M_H=$ 400 GeV (black
long dashed), 500 GeV (black short dashed) and 800 GeV (black
solid). For each curve, the region above the curve is excluded.}
\label{EL-C_t-rho_H}
\end{minipage}
\end{center}
\end{figure}

Taking the same approach as in Ref.\,\cite{KRX-PRD2014}, we
calculate the exclusion bound (with vanishing dim-6 ECC ) in the
$C_t$-$\rho_H$ plane for the cases of $M_H=$ 400 GeV, 500 GeV and
800 GeV. The results are plotted in Fig. \ref{EL-C_t-rho_H}. The
region above each curve is the excluded region.

However, as we showed in Ref.\,\cite{KRX-PRD2014} that the
contribution of the dim-6 interaction with large enough $\rho^{}_H
f^{}_W/\Lambda^2$ and/or $\rho^{}_H f^{}_{WW}/\Lambda^2$ may
cancel a part of the dim-4 interaction contribution to make $H$
easier to escape from being excluded by EB than what is shown in
Fig.\,\ref{EL-C_t-rho_H}. Therefore, we should further take into
account the contribution of the dim-6 interaction. Now we have to
deal with all the four parameters. Of course it is not judicious
to plot a four dimensional figure. Note that we are mainly aiming
at analyzing the most sensitive process VH$^\ast$ which is
actually not sensitive to $C^{}_t$ ($C^{}_t$ only affect the total
width of $H$). So we can simply take $C^{}_t=1$ to represent the
Type-I case, and take $C^{}_t=0.1$ to represent the Type-II case.
It is still not easy to read out the exclusion bound
quantitatively from a three dimensional plot. So we still need to
reduce one parameter. Note that the detection of $H$ from the
VH$^\ast$ process needs a not so small $\rho^{}_H$. So that the
range of $\rho^{}_H$ we are considering is not large. Therefore we
can take $\rho^{}_H=0.2,~0.6$ and 1 to represent three small
regions of $\rho^{}_H$. Then we can plot a two dimensional
exclusion bound in the $\rho^{}_H f^{}_W/\Lambda^2$-$\rho^{}_H
f^{}_{WW}/\Lambda^2$ plane which can be quantitatively read. The
values of the four parameters we are taking are listed in
Table\,\ref{samples}.
\begin{center}
\begin{table}[h]                           
\caption{\label{samples}Values of $C^{}_t$ and $\rho^{}_H$ in our
study.}
\begin{tabular}{c|cc|cc|cc}
\hline\hline
 Label       &A    &        &B    &        &C    &        \\
 Parameter   &$C_t$&$\rho_H$&$C_t$&$\rho_H$&$C_t$&$\rho_H$\\
 \hline
 400GeV~I       &1  &0.2&1  &0.6&1  &1  \\
 500GeV~I       &1  &0.2&1  &0.6&1  &1  \\
 800GeV~I       &1  &0.2&1  &0.6&1  &1  \\
 \hline
 400GeV~II      &0.1&0.2&0.1&0.6&0.1&1  \\
 500GeV~II      &0.1&0.2&0.1&0.6&0.1&1  \\
 800GeV~II      &0.1&0.2&0.1&0.6&0.1&1  \\
 \hline\hline
\end{tabular}
\vspace{5pt}
\end{table}
\end{center}

\begin{widetext}

\begin{figure}[h]
\begin{center}
\begin{minipage}{1\textwidth}
\centering
\includegraphics[width=11cm,height=7cm]{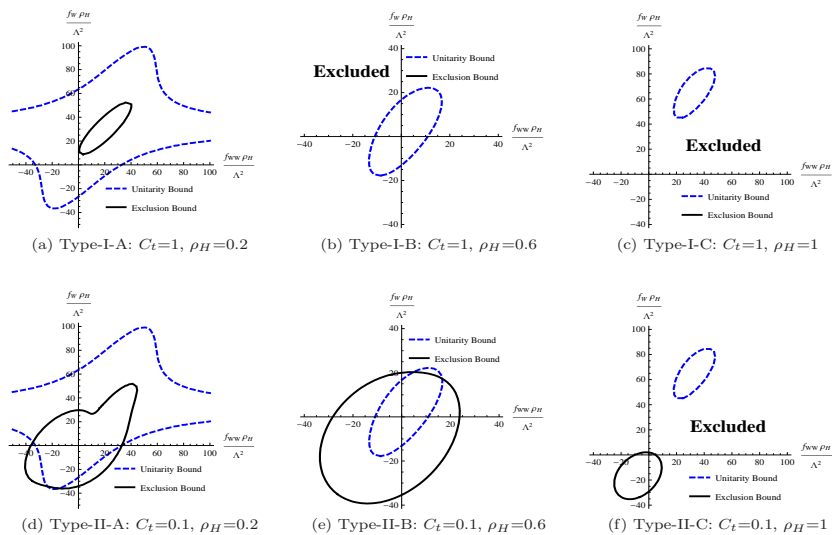}  
\vspace{5pt} \caption{\label{EL-UB-400} Exclusion bounds (outside
the dark-solid contour) and the unitary bound (outside the
blue-dashed contour) for $M_H=$ 400 GeV in the $\rho^{}_H f^{}_W
/\Lambda^2$-$\rho^{}_H f_{WW}/\Lambda^2$ plane (in TeV$^{-2}$).}
\end{minipage}
\end{center}
\end{figure}
\end{widetext}

Taking again the same approach as in Ref.\,\cite{KRX-PRD2014}, we
obtain the exclusion bounds for $M^{}_H=400$ GeV
(Fig.\,\ref{EL-UB-400}), $M^{}_H=500$ GeV (Fig.\,\ref{EL-UB-500}),
and $M^{}_H=800$ GeV (Fig.\,\ref{EL-UB-800}). In these figures,
the region inside the dark-solid contour is not excluded, and the
blue-dashed curves denote the UB.
\begin{widetext}

\begin{figure}[h]
\includegraphics[width=11cm,height=7cm]{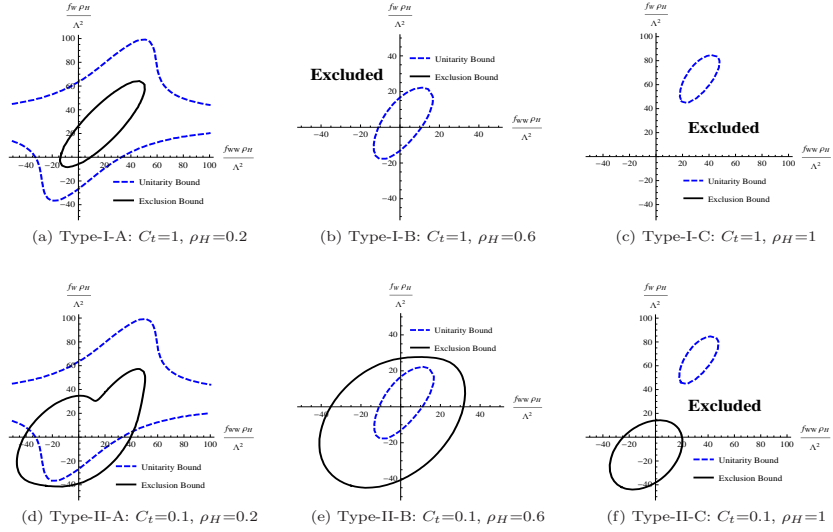}
\vspace{5pt} \caption{\label{EL-UB-500} Exclusion bounds (outside
the dark-solid contour) and the unitary bound (outside the
blue-dashed contour) for $M_H=$ 500 GeV. in the $\rho^{}_H f^{}_W
/\Lambda^2$-$\rho^{}_H f_{WW}/\Lambda^2$ plane (in TeV$^{-2}$). }
\end{figure}

\begin{figure}[h]
\includegraphics[width=11cm,height=7cm]{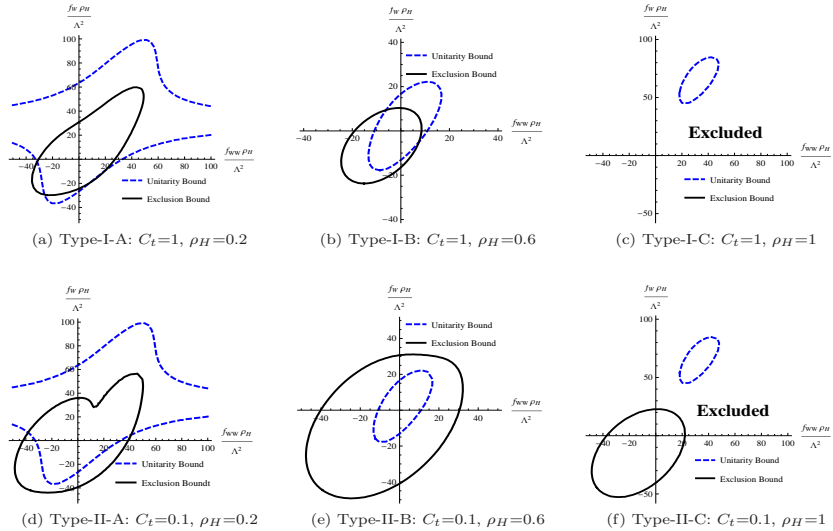}
\vspace{5pt} \caption{\label{EL-UB-800} Exclusion bounds (outside
the dark-solid contour) and the unitary bound (outside the
blue-dashed contour) for $M_H=$ 800 GeV. in the $\rho^{}_H f^{}_W
/\Lambda^2$-$\rho^{}_H f_{WW}/\Lambda^2$ plane (in TeV$^{-2}$).}
\end{figure}

\end{widetext}

 Figure without a dark-solid
contour means that the whole region of $\rho^{}_H
f^{}_W/\Lambda^2$ and $\rho^{}_H f^{}_{WW}/\Lambda^2$ is
excluded.(e.g., the cases of Type-I-B (Fig.\,\ref{EL-UB-400}(b)
and Fig.\,\ref{EL-UB-500}(b)), Type-I-C (Fig.\,\ref{EL-UB-400} (c)
and Fig.\,\ref{EL-UB-500}(c)) for $M^{}_H=400$ and 500 GeV, and
Type-I-C (Fig.\,\ref{EL-UB-800} (c)) for $M^{}_H=800$ GeV. In the
cases of Type-II-C (Fig.\,\ref{EL-UB-400} (f),
Fig.\,\ref{EL-UB-500} (f), and Fig.\,\ref{EL-UB-800} (f)) for
$M^{}_H=400,~500$ and 800 GeV, even there are dark-solid contours,
but they do not overlap with the blue-dashed contours of UB, so
that they are also completely excluded. Thus there are only ten
parameter sets not being excluded which should be considered in
the following sections, namely Type-I-A, Type-II-A, Type-II-B for
$M^{}_H=400$ and 500 GeV (Fig.\,\ref{EL-UB-400} (a), (d), (e)),
Fig.\,\ref{EL-UB-500} (a), (d), (e)), and Type-I-A, Type-I-B,
Type-II-A, Type-II-B (Fig.\,\ref{EL-UB-800} (a), (b), (d), (e))
for $M^{}_H=800$ GeV.

We see that the parameter set $C^{}_t=1,~\rho^{}_H=0.2$ for
$M^{}_H=400$ GeV is in the excluded regions in
Fig.\,\ref{EL-C_t-rho_H}. However, Fig.\,\ref{EL-UB-400} (a) shows
that there is still a region inside the dark-solid contours not
excluded. This means Fig.\,\ref{EL-C_t-rho_H} (neglecting the
dim-6 interactions) is too crude, and dim-6 interactions have to
be taken into account.

\section{Analysis of VH$^\ast$ at LHC Run 2}
\begin{figure}[h]
\begin{center}
\begin{minipage}{0.5\textwidth}
\vspace{5pt}
\includegraphics[width=0.7\textwidth]{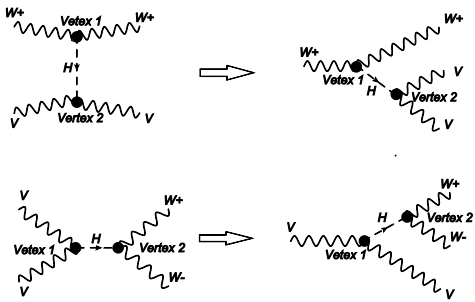}      
\vspace{5pt} \caption{\label{diagrams}Feynman diagrams showing the
relation between WBS (left) and VH$^\ast$ (right).}
\end{minipage}
\end{center}
\end{figure}

In Ref.\,\cite{KRX-PRD2014}, we proposed that the semileptonic
modes of WBS
 and VH$^\ast$
are two sensitive processes for discovering $H$ and measuring its
dim-6 interactions at the 14 TeV LHC. The typical Feynman diagrams
for WBS and VH$^\ast$ (having crossing symmetry) with the same ECC
and the relation between them are shown in Fig. \ref{diagrams}. So
their sensitivity of depending on the ECC $\rho^{}_H f^{}_W
/\Lambda^2$ and $\rho^{}_H f^{}_{WW}/\Lambda^2$ (in the dim-6
interaction) should be similar. Since the most sensitive process
is VH$^\ast$, we concentrate on analyzing the VH$^*$ process in
this section. We shall calculate the the ranges of $\rho^{}_H
f^{}_W /\Lambda^2$ and $\rho^{}_H f^{}_{WW}/\Lambda^2$
corresponding to the $1\sigma,~3\sigma$ and $5\sigma$ statistical
significance for the ten allowed parameter sets of $C^{}_t$ and
$\rho^{}_H$ memtioned in Sec.\,2 for an integrated luminosity of
100 fb$^{-1}$ at the 14 TeV LHC.

We use MadGraph5 \cite{madgraph5} interfaced with FeynRules
\cite{feynrules} and Pythia6.4 \cite{pythia-6.4} to simulate
signals and backgrounds, and take CTEQ6.1 \cite{cteq6.1} as the
parton distribution function (PDF). Delphes3 \cite{delphes3} and
fastjet \cite{fastjet} is used to simulate detector acceptance and
jet reconstruction. The detector acceptance is set in Table
\ref{acceptance} referring to the design of CMS detector
\cite{CMS-detector}.

\begin{table}[h]
\begin{center}
\caption{\label{acceptance}Detector acceptance according to
DELPHES3}
\begin{tabular}{ccccc}
\hline\hline
 &$\mu$&$e$&{\rm jet}&{\rm photon}\\
 \hline
 $|\eta|_{max}$~~~~~~~~~~&2.4&2.5&5 &2.5\\
 $p^{}_{Tmax} $(GeV)     &10 &10 &20&0.5\\
 \hline\hline
\end{tabular}
\end{center}
\end{table}

We use the Cambridge/Aachen (C/A) algorithm with radius $R$=0.8
\cite{ca-jet} to cluster the boosted jets and then apply jet
pruning algorithm \cite{prune} with parameters Z$_{\rm cut}$=0.1
and RFactor$_{\rm cut}$=0.5 on the C/A jets. Then we apply the
same cuts as in Ref.\,\cite{KRX-PRD2014}. In addition, we only
take the events within a small vicinity around the resonance peak
of $H$ as what we did in Ref.\,\cite{KRX-PRD2014}. The jet pruning
algorithm further suppresses the backgrounds.

Let $\sigma^{}_S$ and $\sigma^{}_B$ be the cross sections of the
signal and background, respectively. For an integrated luminosity
${\cal L}^{}_{\rm int}$, the event numbers $N^{}_S$ and $N^{}_B$
of the signal and background are $N^{}_S={\cal L}^{}_{\rm
int}\sigma^{}_S$ and $N^{}_B={\cal L}^{}_{\rm int}\sigma^{}_B$. In
the case of ${\cal L}^{}_{\rm int}=$ 100 fb$^{-1}$ at the 14 TeV
LHC, $N^{}_S$ and $N^{}_B$ are large, so that the statistical
significance $\sigma^{}_{\rm stat}$ can be approximately expressed
as
\begin{equation}                               
\label{significance} \sigma_{\rm
stat}=\frac{N^{}_{S}}{\sqrt{N^{}_{B}}}.
\end{equation}

\begin{widetext}

\begin{figure}[h]
\begin{center}
\begin{minipage}{1\textwidth}
\centering
\includegraphics[width=11cm,height=3.5cm]{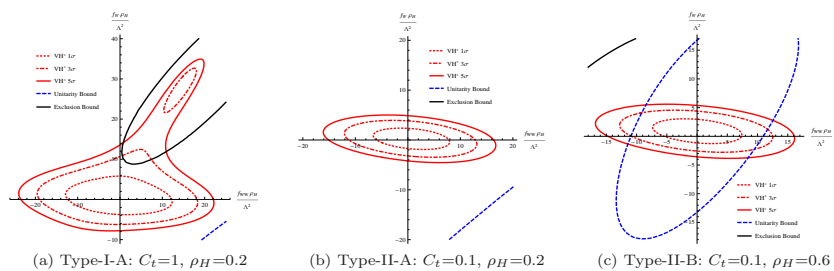} 
\vspace{5pt} \caption{\label{VH-400} Contours for 1$\sigma$,
3$\sigma$ and 5$\sigma$ statistical significance for VH$^\ast$ on
the  $\rho^{}_H f^{}_W /\Lambda^2$-$\rho^{}_H f^{}_{WW}/\Lambda^2$
plane (in TeV$^{-2}$) for $M_H=$ 400 GeV at the 14 TeV LHC with
${\cal L}^{}_{\rm int}=100$ fb$^{-1}$. The EB (dark-solid) and/or
UB (blue-dashed) are also shown (or partly shown).}
\end{minipage}
\end{center}
\end{figure}
\begin{figure}[h]
\includegraphics[width=11cm,height=3.5cm]{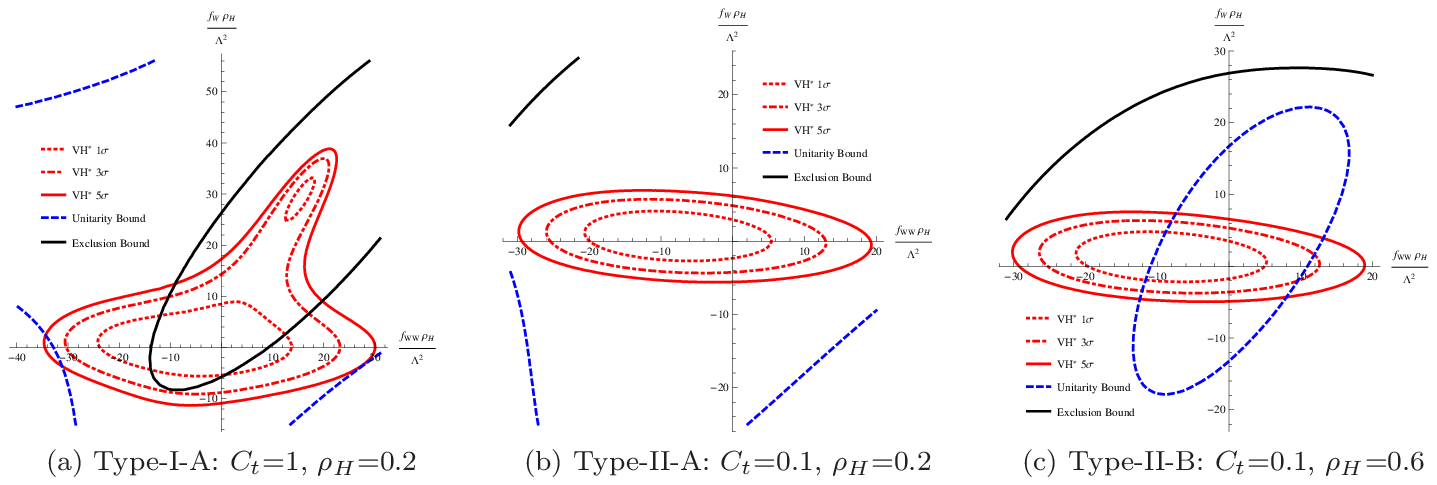} \vspace{5pt}
\caption{\label{VH-500} Contours for 1$\sigma$, 3$\sigma$ and
5$\sigma$ statistical significance for VH$^\ast$ on the $\rho^{}_H
f^{}_W /\Lambda^2$-$\rho^{}_H f^{}_{WW}/\Lambda^2$ plane (in
TeV$^{-2}$) for $M_H=$ 500 GeV at the 14 TeV LHC with ${\cal
L}^{}_{\rm int}=100$ fb$^{-1}$. The EB (dark-solid) and/or UB
(blue-dashed) are also shown (or partly shown).}
\end{figure}
\begin{figure}[h]
\includegraphics[width=9cm,height=9.2cm]{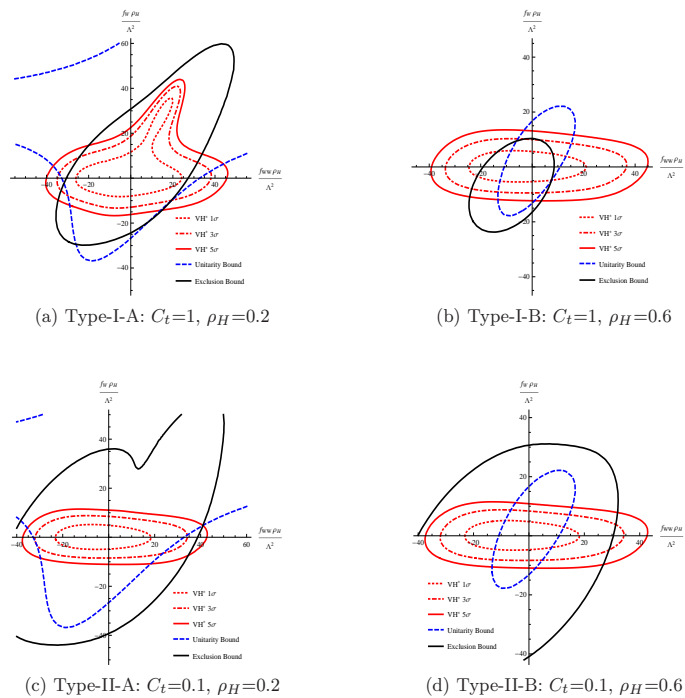}
\vspace{5pt} \caption{\label{VH-800} Contours for 1$\sigma$,
3$\sigma$ and 5$\sigma$ statistical significance for VH$^\ast$ on
the  $\rho^{}_H f^{}_W /\Lambda^2$-$\rho^{}_H f^{}_{WW}/\Lambda^2$
plane (in TeV$^{-2}$) for $M_H=$ 800 GeV at the 14 TeV LHC with
${\cal L}^{}_{\rm int}=100$ fb$^{-1}$. The EB (dark-solid) and/or
UB (blue-dashed) are also shown (or partly shown).}
\end{figure}

\end{widetext}

In Fig.\,\ref{VH-400}, Fig.\,\ref{VH-500}, and Fig.\,\ref{VH-800},
we plot the contours (red-dotted, red-dashed-dotted, and
red-solid), in the $\rho^{}_H f^{}_W /\Lambda^2$-$\rho^{}_H
f^{}_{WW}/\Lambda^2$ plane, corresponding to the statistical
significance of 1$\sigma$ (margin), 3$\sigma$ (evidence) and
5$\sigma$ (discovery) for the process VH$\ast$ with $M^{}_H=$ 400,
500, and 800 GeV, respectively. In these figures, we also plot (or
partly plot) the EB (dark-solid) and/or the UB (blue-dashed) given
in Sec.\,2 to show the actual allowed regions. The ten figures in
Fig.\,\ref{VH-400}, Fig.\,\ref{VH-500}, and Fig.\,\ref{VH-800} are
for the ten sets of $C^{}_t$ and $\rho^{}_H$ mentioned in Sec.\.2.

we see that, in most cases, EB and UB put nontrivial constraints
on the red contours. Only some parts of the red contours inside
the allowed regions set by the EB and/or UB are actually allowed,
while the parts outside the allowed regions are excluded. The only
exception is the case of Type-II-A for $M^{}_H=$ 400 GeV whose red
contours are so small that they are completely well within the
allowed region.

In the following, we discuss two useful applications of these
results.

\null\noindent{\bf (A) Experimental determination of
$\bm{\rho^{}_H f^{}_W /\Lambda^2}$ and $\bm{\rho^{}_H
f^{}_{WW}/\Lambda^2}$ of $\bm H$}

In Ref.\,\cite{KRX-PRD2014}, we pointed  out that, after the
discovery of the resonance peak of $H$, one can further measure
four distributions, namely the $p^{}_T(\rm leptons)$-, the
$p^{}_T(J^{}_1)$-, the $\Delta R(\ell^+,J^{}_1)$-, and the $\Delta
R(J^{}_1,J^{}_2)$-distribution, to determine of values of
$\rho^{}_H f^{}_W /\Lambda^2$ and $\rho^{}_H f^{}_{WW}/\Lambda^2$
of this $H$ (cf. Sec.\,VIII of Ref.\,\cite{KRX-PRD2014}). Now we
can see the specific way of realizing it from Fig.\,\ref{VH-400},
Fig.\,\ref{VH-500}, and Fig.\,\ref{VH-800}. Taking the $5\sigma$
discovery of $H$ in the the case of Type-II-B for $M^{}_H=$ 500
GeV (Fig.\,\ref{VH-500} (c)) as an example, the allowed values of
$\rho^{}_H f^{}_W /\Lambda^2$ and $\rho^{}_H f^{}_{WW}/\Lambda^2$
lie on two segments of the red-solid contour inside the UB allowed
region. Thus we can determine the values $\rho^{}_H f^{}_W
/\Lambda^2$ and $\rho^{}_H f^{}_{WW}/\Lambda^2$ by adjusting the
values on these two segments in the theoretical distributions to
fit the experimentally measured distributions. Since these two
segments are not long, the best fit values may be easily obtained
by iteration. 
The so determined
values of $\rho^{}_H f^{}_W /\Lambda^2$ and $\rho^{}_H
f^{}_{WW}/\Lambda^2$ serve as {\it a new powerful high energy
criterion for discriminating new physics models}. {\it Only models
whose predicted $\rho^{}_H f^{}_W /\Lambda^2$ and $\rho^{}_H
f^{}_{WW}/\Lambda^2$ are consistent with the experimentally
determined values can survive as candidates of the correct new
physics models reflecting the nature. All models whose predicted
$\rho^{}_H f^{}_W /\Lambda^2$ and $\rho^{}_H f^{}_{WW}/\Lambda^2$
are not consistent with the experimentally determined values will
be ruled out.}

\null\noindent{\bf (B) Theoretical exclusion bounds if $H$ is not
discovered at LHC Run 2}\\
\null~~~~ In this paper, we take into account only the statistical
error, and leave the study of the systematic error to
experimentalists. In this sense, the $1\sigma$ contours for the
ten possible parameter sets (cf. Sec.\,2) shown in
Figs.\,\ref{VH-400}, \ref{VH-500}, and \ref{VH-800} play a
important role. For each set of $C^{}_t$ and $\rho^{}_H$, the
regions inside the $1\sigma$ contour means that the signal is
immersed in the statistical fluctuation, i.e., it cannot be
detected. Thus, theoretically, if the resonance peak is not found
at the 14 TeV LHC, {\it the $1\sigma$ contours provide the
strongest theoretical exclusion bound on $\rho^{}_H f^{}_W
/\Lambda^2$ and $\rho^{}_H f^{}_{WW}/\Lambda^2$ for each set of
$C^{}_t$ and $\rho^{}_H$, i.e., the values of $\rho^{}_H f^{}_W
/\Lambda^2$ and $\rho^{}_H f^{}_{WW}/\Lambda^2$ outside the
$1\sigma$ contours are excluded}. Note that in Fig.\,\ref{VH-400}
(a) the $1\sigma$ contour is completely in the excluded region.
{\it In this case, the whole allowed region is excluded}.

\section{Analysis of GF and VBF at LHC Run 2}

On-shell Higgs productions via GF and VBF are traditional
processes in the discovery and measurement of the 125 GeV Higgs
boson $h$ at the LHC Run 1. The most accurate measurement comes
from the decay mode $h\to ZZ\to 4\ell$. In
Ref.\,\cite{KRX-PRD2014}, we pointed out that the dim-6
interactions are suppressed by a factor $k^2/\Lambda^2$ relative
to the dim-4 interactions, where $k$ is a typical momentum scale
(from the extra derivatives in the dim-6 interactions) appearing
in the dim-6 interaction, and it is of the order of the momentum
of the Higgs boson. In on-shell Higgs productions of the heavy
Higgs boson $H$, $k^2\sim M^2_H$. 
Taking $M^{}_H=500$ GeV with $\Lambda=3$ TeV as an example,
$k^2/\Lambda^2\sim (500/3000)^2=0.03$. This means that the dim-6
interactions only contribute about $3\%$ of the total
contribution. So that it is hard to measure the effect of the
dim-6 interactions in on-shell Higgs productions. This is the
reason why we concentrate our study on the VH$^\ast$ process.
However, in certain regions of the ECC, on-shell productions of
$H$ via GF and VBF may still help for discovering $H$. So, for
completeness, we analyze these two processes in this section.

The signals and backgrounds for the GF and VBF processes in the
LHC Run 1 have been analyzed in Ref.\,\cite{CMS-higgs-4l}. Here we
take the same approach as in Sec.\,2. For the signals, we take the
production cross sections and branching ratios given by the LHC
Higgs Cross Section Working Group \cite{Higgs-cross-section} and
rescale their distributions. For the main background of GF, $pp
\rightarrow ZZ \rightarrow 4\ell$, we rescale it with the K-factor
given in Ref.\,\cite{pp-zz-NLO}. We take the anti-k$_T$ algorithm
with radius $R$=0.5 \cite{anti-kt-jet} to cluster jets and refer
to the research of the CMS collaboration on $4\ell$ mode of Higgs
decay \cite{CMS-higgs-4l} to apply cuts in this section. The
events in which the final four leptons can reconstruct the mass of
$H$ are selected for both the signal and the background processes.

\begin{widetext}

\begin{figure}[h]
\begin{center}
\begin{minipage}{1\textwidth}
\centering
\includegraphics[width=12cm,height=3cm]{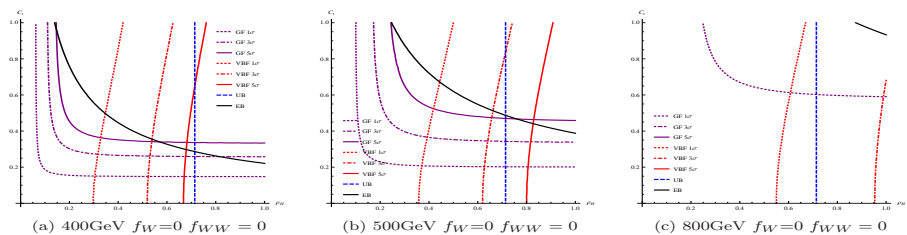} 
\vspace{5pt} \caption{\label{VBF-GF-C_t-rho_H} The 1$\sigma$
(dotted), 3$\sigma$ (dashed-dotted) and 5$\sigma$ (solid) contours
of the GF (purple) and VBF (red) processes for $H$ with vanishing
dim-6 ECC at the 14 TeV LHC with ${\cal L}^{}_{\rm int}=100$
fb$^{-1}$. The EB (dark-solid, from Fig.\,\ref{EL-C_t-rho_H}) and
UB (blue-dotted) are also shown.}
\end{minipage}
\end{center}
\end{figure}

\end{widetext}

Since the dim-4 interaction dominates in these two on-shell $H$
production processes, we first analyze it neglecting the dim-6
interactions. The $1\sigma,~3\sigma$ and $5\sigma$ contours with
vanishing dim-6 ECC are plotted in Fig.\,\ref{VBF-GF-C_t-rho_H}.

We see from Fig.\,\ref{VBF-GF-C_t-rho_H} that GF is sensitive for
discovering $H$ when $C^{}_t$ and $\rho^{}_H$ are both not so
small. However, as we see from Fig.\,\ref{VBF-GF-C_t-rho_H}, quite
a large portion of this region has already been excluded by EB.
The VBF process is sensitive when $\rho_H$ is large, but UB
excludes the $5\sigma$ discovery for $M^{}_H>$ 500 GeV, and allows
a very narrow region for $5\sigma$ discovery only for $M^{}_H=$
400 GeV.


\begin{widetext}

\begin{figure}[h]
\begin{center}
\begin{minipage}{1\textwidth}
\centering
\includegraphics[width=13cm,height=5cm]{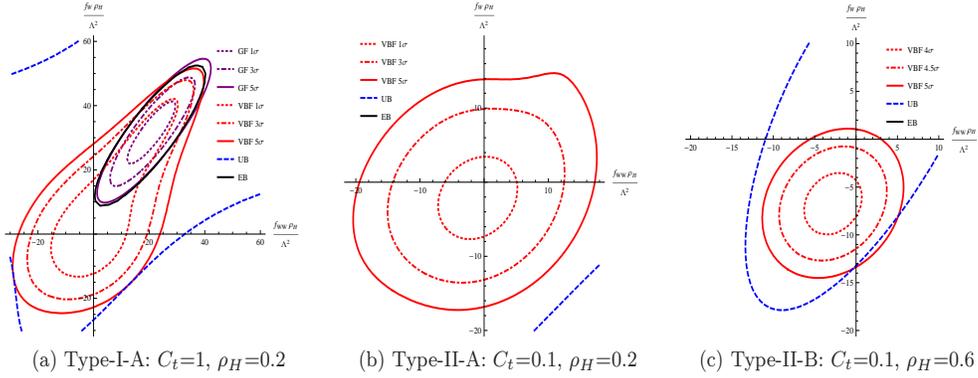}  
\vspace{5pt} \caption{\label{VBF-GF-400} $1\sigma,~3\sigma$ and
$5\sigma$ contours for GF (purple-dotted, purple-dashed-dotted,
and purple-solid)) and VBF (red-dotted, red-dashed-dotted, and
red-solid) in the $\rho^{}_H f^{}_W /\Lambda^2$-$\rho^{}_H
f^{}_{WW}/\Lambda^2$ plane (in TeV$^{-2}$) for $M_H=$ 400 GeV at
the 14 TeV LHC with ${\cal L}^{}_{\rm int}=$ 100 fb$^{-1}$. Except
in (a), the tiny contours for GF are ignored. }
\end{minipage}
\end{center}
\end{figure}
\begin{figure}[h]
\includegraphics[width=13cm,height=5cm]{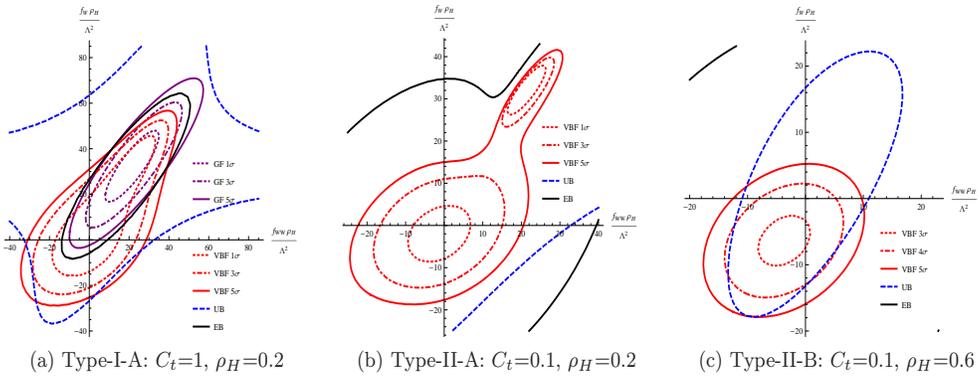}
\vspace{5pt} \caption{\label{VBF-GF-500}  $1\sigma,~3\sigma$ and
$5\sigma$ contours for GF (purple-dotted, purple-dashed-dotted,
and purple-solid)) and VBF (red-dotted, red-dashed-dotted, and
red-solid) in the $\rho^{}_H f^{}_W /\Lambda^2$-$\rho^{}_H
f^{}_{WW}/\Lambda^2$ plane (in TeV$^{-2}$) for $M_H=$ 500 GeV at
the 14 TeV LHC with ${\cal L}^{}_{\rm int}=$ 100 fb$^{-1}$. Except
in (a), the tiny contours for GF are ignored.}
\end{figure}
\end{widetext}

\begin{widetext}

\begin{figure}[h]
\includegraphics[width=10cm,height=9cm]{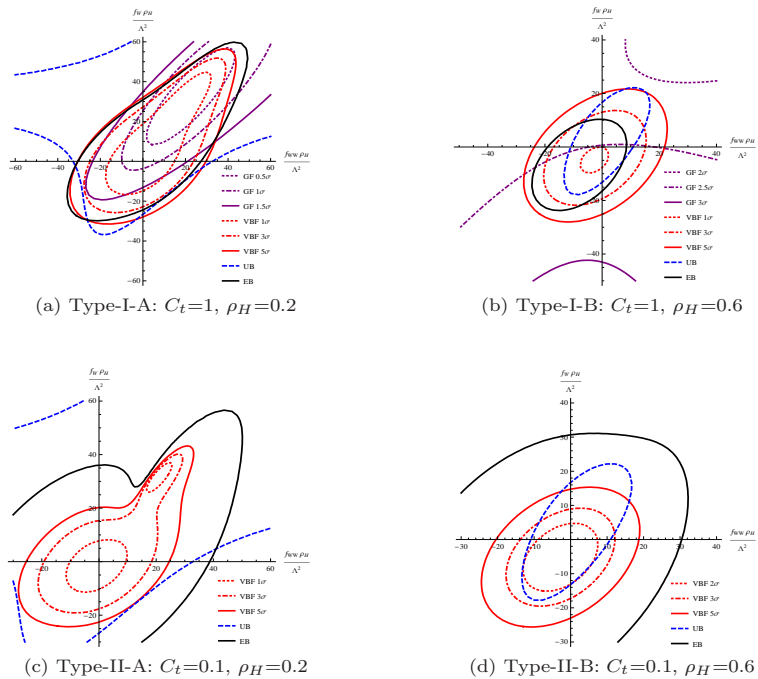}
\vspace{5pt} \caption{\label{VBF-GF-800}  $1\sigma,~3\sigma$ and
$5\sigma$ contours for GF (purple-dotted, purple-dashed-dotted,
and purple-solid)) and VBF (red-dotted, red-dashed-dotted, and
red-solid) in the $\rho^{}_H f^{}_W /\Lambda^2$-$\rho^{}_H
f^{}_{WW}/\Lambda^2$ plane (in TeV$^{-2}$) for $M_H=$ 800 GeV at
the 14 TeV LHC with ${\cal L}^{}_{\rm int}=$ 100 fb$^{-1}$. Except
in (a) and (b), the tiny contours for GF are ignored.}
\end{figure}
\end{widetext}

Next we analyze the general case including the dim-6 interactions.
The $1\sigma$, $3\sigma$ and $5\sigma$ contours for GF (purple)
and VBF (red) together with the EB (dark-solid) and UB
(blue-dashed) constraints for $M^{}_H=$ 400, 500, and 800 GeV are
plotted in Fig.\,\ref{VBF-GF-400}, \ref{VBF-GF-500} and
\ref{VBF-GF-800}, respectively for the ten sets of $C^{}_t$ and
$\rho^{}_H$ mentioned in Sec.\,2.

We see that GF can help to discover $H$ only in the case of
Type-I-A with very narrow available parameter ranges, and can
hardly discover $H$ in all other cases. VBF can help to discover
$H$ in more cases except Type-I-B for $M^{}_H=800$ GeV, but the
available parameter ranges are all quite small.

 Comparing Fig.\,\ref{VBF-GF-400}
(a) with Fig.\,\ref{VH-400} (a), we see that the $1\sigma$ contour
for GF and VBF are larger than that for VH$^\ast$. So that if $H$
is not discovered, VH$^\ast$ still gives the strongest exclusion
bound.

We also see that the density of the contours for VH$^*$ process is
much larger than that for GF and VBF.  This means that VH$^\ast$
is much more sensitive to the variation of $\rho^{}_H f^{}_W
/\Lambda^2$ and $\rho^{}_H f^{}_{WW}/\Lambda^2$. This is why we
only suggest measuring $\rho^{}_H f^{}_W /\Lambda^2$ and
$\rho^{}_H f^{}_{WW}/\Lambda^2$ via VH$^\ast$.

\section{Summary}

In this paper, we extend the study in Ref.\,\cite{KRX-PRD2014} to
a more thorough analysis of EB from the LHC Run 1 data, the UB,
and the relations between the statistical significance oof
$1\sigma,~3\sigma,~5\sigma$ and the ranges of the ECC in the
general effective interactions related to the heavy neutral Higgs
boson $H$. These results are very useful in the Run 2 of the LHC
for realizing the experimental determination of $\rho^{}_H f^{}_W
/\Lambda^2$ and $\rho^{}_H f^{}_{WW}/\Lambda^2$ if $H$ is
discovered, and setting the exclusion bounds on the ECC if $H$ is
not found.

We take the same formulation of the effective interactions related
to the heavy neutral Higgs boson $H$ as in
Ref.\,\cite{KRX-PRD2014}, which contains five parameters, namely
the heavy Higgs mass $M^{}_H$, the anomalous $Ht\bar t$ Yukawa
coupling factor $C^{}_t$, the anomalous gauge coupling constant
$\rho^{}_H$ in the dim-4 HVV interactions, and the anomalous gauge
coupling constants $\rho^{}_H f^{}_W /\Lambda^2$ and $\rho^{}_H
f^{}_{WW}/\Lambda^2$ in the dim-6 HVV interactions. We take
$M^{}_H=400$ GeV, 500 GeV, and 800 GeV to represent three mass
ranges of $M^{}_H$ in this study.

It has been pointed out that, at the 14 TeV LHC, the most
sensitive processes for discovering $H$ and measuring its
$\rho^{}_H f^{}_W /\Lambda^2$ and $\rho^{}_H f^{}_{WW}/\Lambda^2$
is $pp\to VH^\ast\to VVV$ (VH$^\ast$). So we concentrate on
analyzing the process VH$^\ast$ in this paper. Since VH$^\ast$ is
not sensitive to the variation of $C^{}_t$, we just take two
values of $C^{}_t$, namely $C^{}_t=1$ and $C^{}_t=0.2$ to
represent the two types of anomalous Yukawa interactions, Type-I
and Type-II, respectively. In addition, the process VH$^\ast$ is
detectable only if the HVV interactions are not so weak (the probe
of heavy Higgs bosons with very weak HVV interactions
(gauge-phobic or nearly gauge-phobic) is given in
Ref.\,\cite{KX-PLB2015}), so we consider a not so large range of
$\rho^{}_H$, namely $0.2<\rho^{}_H<1$, and divide it to three
parts. We take $\rho^{}_H= 0.2,~0.6$ and 1 to represent these
three parts. This parameter setting of $C^{}_t$ and $\rho^{}_H$ is
shown in Tab.\,\ref{samples}.

We first gave a more thorough study of the EB from the LHC Run 1
data, and the UB from the requirement of the unitarity of the $S$
matrix elements in Sec. 2 for the parameter sets given in
Tab.\,\ref{samples}. This already gives quite strong constraints
on the ECC, and we shall see it plays an important role in the
analysis of the VH$^\ast$ in Sec.\,3.

Sec.\,3 is the main part of our analysis. We calculated the
contours for the statistical significance of $1\sigma$ (margin),
$3\sigma$ (evidence), and $5\sigma$ (discovery) with the
integrated luminosity ${\cal L}^{}_{\rm int}=100$ fb$^{-1}$ for
the process VH$^\ast$ at the 14 TeV LHC. The results are plotted
in Figs.\,\ref{VH-400},\,\ref{VH-500}, and \ref{VH-800}. These
results has two useful applications in the Run 2 of the LHC: (A)
realizing the experimental determination of $\rho^{}_H f^{}_W
/\Lambda^2$ and $\rho^{}_H f^{}_{WW}/\Lambda^2$ which provides a
new high energy criterion for discriminating new physics models,
i.e., {\it Only models whose predicted $\rho^{}_H f^{}_W
/\Lambda^2$ and $\rho^{}_H f^{}_{WW}/\Lambda^2$ are consistent
with the experimentally determined values can survive as
candidates of the correct new physics models reflecting the
nature.}, (B) Setting the exclusion bounds on the ECC from the
$1\sigma$ contours if $H$ is not found at the LHC Run 2. These are
important extensions of the study in Ref.\,\cite{KRX-PRD2014}.

Finally, for completeness, we also analyzed the traditional
processes of on-shell Higgs productions via GF and VBF in Sec.\,
4. the results are shown in Figs.\,\ref{VBF-GF-C_t-rho_H},
\ref{VBF-GF-400}, \ref{VBF-GF-500}, and \ref{VBF-GF-800}. First of
all, we showed that on-shell Higgs productions via GF and VBF can
hardly give contribution to the experimental determination of
$\rho^{}_H f^{}_W /\Lambda^2$ and $\rho^{}_H f^{}_{WW}/\Lambda^2$.
Then from Figs.\,\ref{VBF-GF-C_t-rho_H}, \ref{VBF-GF-400},
\ref{VBF-GF-500}, and \ref{VBF-GF-800} we see that: (i) GF can
help to discover $H$ only in the case of Type-I-A with very narrow
available parameter ranges, and can hardly discover $H$ in all
other cases; (ii) VBF can help to discover $H$ in more cases
except Type-I-B for $M^{}_H=800$ GeV, but the available parameter
ranges are all quite small; (iii) if $H$ is not found at the LHC
Run 2 experiments, the exclusion bounds on ECC from GF and VBF are
significantly weaker than those from VH$^\ast$.

In a word, we conclude that VH$^\ast$ is the best process for
discovering $H$ and measuring its $\rho^{}_H f^{}_W /\Lambda^2$
and $\rho^{}_H f^{}_{WW}/\Lambda^2$ at Run 2 of the LHC.

\mbox{}

\acknowledgments{We are grateful to Guo-Ming Chen for valuable
discussions. We would like to thank Tsinghua National Laboratory
for Information Science and Technology for providing their
computing facility. This work is supported by the National Natural
Science Foundation of China under the grant numbers 11135003 and
11275102.}

\vspace{10mm}

\vspace{-1mm} \centerline{\rule{80mm}{0.1pt}} \vspace{2mm}


\begin{thebibliography}{99}
\bibitem{footnote}
This theoretical exclusion bound is not the same as the
experimental $95\%$ C.L. exlcusion bound which depends on the data
and the total error (including the systematic error). The
experimental $95\%$ C.L. exlcusion bound can only be obtained when
the measurement is done, and can only be given by the
experimentalists.
\end{thebibliography}

\clearpage

\end{document}